\documentclass[11pt]{article}

\usepackage[utf8]{inputenc}
\usepackage[T1]{fontenc}
\usepackage{lmodern}
\usepackage{microtype}
\usepackage{geometry}
\usepackage{amsmath,amssymb,amsthm}
\usepackage{booktabs}
\usepackage{array}
\usepackage{natbib}
\usepackage[hidelinks]{hyperref}
\hypersetup{pdfauthor={Christian Jaag},pdftitle={Competition with a Common Purpose}}
\usepackage{setspace}
\usepackage{enumitem}
\usepackage{mathtools}

\newtheorem{proposition}{Proposition}

\theoremstyle{remark}

\title{Competition with a Common Purpose}
\author{Christian Jaag\thanks{Swiss Economics, Ottikerstrasse 7, 8006 Zürich, Switzerland and Somaha Foundation, Weinbergstrasse 102, 8006 Zürich, Switzerland. Corresponding author: \href{mailto:christian.jaag@somaha-stiftung.ch}{christian.jaag@somaha-stiftung.ch}.}}
\date{September 2026}

\begin{document}

\maketitle

\begin{abstract}
Non-governmental organizations often compete for funding while valuing similar social outcomes. Fundraising may expand total giving to a cause or redirect a given funding pool toward the focal organization. This paper studies how cross-organizational purpose alignment changes the composition of such competition. The analysis separates two effects of fundraising: marginal \emph{lift} in aggregate funding and marginal \emph{shift} of funding between organizations. Incomplete alignment creates a wedge between an NGO's marginal mission return and the aggregate mission return that is proportional to shift minus lift. When NGOs choose among fundraising technologies, greater purpose alignment shifts effort toward the relatively lift-intensive technology. Relative to the aggregate mission benchmark, incompletely aligned NGOs choose portfolios that are too shift-intensive. A uniform restriction on fundraising may change total effort but cannot generally correct this composition distortion; correcting the portfolio requires changing relative incentives across technologies. A simple microfoundation interprets cause promotion as lift and organization promotion as shift. Productivity differences preserve reallocation when it raises mission output, while organization-specific benefits restore rivalry even under complete alignment. The model therefore predicts how a common purpose changes the form of competition even when its effect on the level of competition is ambiguous.
\end{abstract}

\noindent\textbf{Keywords:} nonprofit competition; NGOs; fundraising; charitable giving; mission; purpose alignment

\noindent\textbf{JEL codes:} D64; H41; L31

\section{Introduction}

Non-governmental organizations often compete with organizations pursuing closely related social objectives. Humanitarian organizations solicit donations for the same emergency. Environmental organizations seek support for similar conservation goals. NGOs concerned with poverty, health, education, or culture approach overlapping groups of foundations, governments, and private donors. Competition for funding may therefore be intense even when disagreement about the underlying social purpose is small.

This setting differs from standard commercial competition in one basic respect. A commercial firm normally does not value output produced by a rival. A purpose-oriented organization may. An NGO seeking to reduce poverty may value poverty reduction achieved by another NGO. A campaign that merely moves one franc from one equally effective NGO to another then contributes little to the common purpose. A campaign that attracts an additional franc to the cause remains valuable even when another organization ultimately receives it.

The distinction shifts attention from the amount of fundraising to its composition. Some activities expand category-level giving by making a cause more salient, educating donors, or bringing new donors into the market. Other activities mainly raise the focal organization's share of a largely given funding pool. Real campaigns combine both effects. The economic question is therefore how purpose alignment changes the relative attractiveness of fundraising technologies that differ in the extent to which they create new funding rather than reallocate existing funding.

The paper shows that purpose alignment changes the form of fundraising competition. To state this result without committing to a particular donor-demand or contest model, any marginal fundraising activity is summarized by \emph{lift}, its contribution to aggregate funding, and \emph{shift}, its contribution to the funding difference between the focal NGO and its rival. An NGO values its own mission output and assigns a weight to mission output produced by the rival. The latter parameter measures cross-organizational purpose alignment. It need not coincide with textual similarity of mission statements: two NGOs may state the same objective while still attaching substantial value to controlling the resources themselves.

The analysis has one main behavioral result. When two fundraising technologies differ in their lift--shift profiles, greater purpose alignment shifts the fundraising portfolio toward the relatively lift-intensive technology. The result does not require a specific donor-demand system, contest technology, or a polar distinction between cause and organization promotion. It follows from a single-crossing property: alignment raises the relative value of funding that benefits the rival and lowers the relative value of funding transferred away from the rival. The effect on total fundraising remains ambiguous.

The same structure yields a result for fundraising coordination. Relative to an aggregate mission benchmark, incompletely aligned NGOs choose fundraising portfolios that are too shift-intensive. A uniform tax, cap, or increase in the cost of fundraising may change total fundraising but generally leaves this portfolio distortion intact. Coordination must instead alter relative incentives across fundraising technologies. An instrument-specific Pigouvian correction then appears as a corollary: relative support for an activity rises with its lift relative to its shift.

A short microfoundation maps the sufficient statistics into familiar fundraising activities. Cause promotion expands a common donation pool and is pure lift in the polar benchmark. Organization promotion changes relative salience and is pure shift. The microfoundation is illustrative rather than necessary for the main result. Two extensions delimit the coordination result. Productivity differences make some reallocation socially valuable because funding may move toward a more effective NGO. Organization-specific benefits, including autonomy, reputation, survival, or private organizational rents, restore organization-centered competition even when contemporaneous purposes are fully aligned.

The contribution is deliberately narrower than a general theory of nonprofit competition. \citet{RoseAckerman1982} studies excessive fundraising. \citet{AldashevVerdier2010} distinguish competition for a fixed donor market from fundraising that expands giving. \citet{AldashevMariniVerdier2014} analyze fundraising coordination among nonprofits, while \citet{AldashevMariniVerdier2020} study informational spillovers and project clustering. \citet{Scharf2014} provides the closest objective-side benchmark by allowing charities to value total public-good provision while placing an additional premium on own provision. \citet{GayleHarrison2026} model and estimate strategic fundraising reactions. \citet{JaagGrantImpactLoan2026} studies the allocation of scarce impact capital between grants and impact loans, \citet{JaagCatalytic2026} examines catalytic capital and commercial crowding-in, and \citet{JaagFutureIntervention2026} analyzes mission commitment under future contracting. The present paper does not claim novelty for any of these ingredients separately. Its contribution is a behavioral and institutional implication of combining them: cross-organizational valuation changes which fundraising technologies NGOs choose, and a distortion in that composition cannot generally be corrected by regulating fundraising only at the aggregate level.

The empirical distinction underlying the model is also observable. \citet{Meer2017}, \citet{Schmitz2021}, and \citet{ScharfSmithOttoniWilhelm2022} distinguish increases in total giving from substitution across charities or time. The model maps such responses into organizational incentives and predicts how the composition of fundraising should vary with purpose alignment. \citet{JaagAdditionality2026} similarly shows that an observed funding gain at one organization need not represent an equal increase in resources available to the underlying purpose. Here the object of analysis is competition for the funding itself.

The remainder of the paper develops the lift--shift model, derives the endogenous portfolio result, characterizes the aggregate mission benchmark and fundraising coordination, and then considers microfoundations, productivity differences, organization-specific value, and empirical implications.

\section{A lift--shift model of fundraising competition}

\subsection{Organizations and purpose alignment}

There are two NGOs, indexed by $i\in\{1,2\}$. NGO $i$ receives funding $D_i$ and converts one unit of funding into $a_i>0$ units of mission output. Its output is
\begin{equation}
q_i=a_iD_i.
\label{eq:output}
\end{equation}

NGO $i$ values its own output and a fraction $\rho_i$ of output produced by NGO $j$:
\begin{equation}
U_i=q_i+\rho_i q_j-C_i,
\label{eq:utility}
\end{equation}
where $\rho_i\in[0,1]$ and $C_i$ denotes the real resource cost of fundraising. At $\rho_i=0$, only own output enters the objective. At $\rho_i=1$, NGO $i$ values an additional unit of mission output equally regardless of which organization produces it. We refer to $\rho_i$ as \emph{purpose alignment}.

The baseline analysis considers identical NGOs, $a_i=a_j=a$ and $\rho_i=\rho_j=\rho$. Heterogeneous productivity is introduced in Section~\ref{sec:extensions}.

\subsection{Lift and shift}

Consider a marginal fundraising action $e_i$. At a symmetric allocation, define its effects on own and rival funding by
\begin{equation}
\frac{\partial D_i}{\partial e_i}=L+S,
\qquad
\frac{\partial D_j}{\partial e_i}=L-S.
\label{eq:LSresponses}
\end{equation}
The sufficient statistic
\begin{equation}
L=\frac12\left(\frac{\partial D_i}{\partial e_i}+\frac{\partial D_j}{\partial e_i}\right)
\end{equation}
is half the marginal change in aggregate funding. We call it \emph{lift}. The statistic
\begin{equation}
S=\frac12\left(\frac{\partial D_i}{\partial e_i}-\frac{\partial D_j}{\partial e_i}\right)
\end{equation}
is half the marginal change in the funding difference and is called \emph{shift}. The decomposition is local and does not impose a donor utility function or a particular contest technology.

For identical productivity $a$, the private marginal mission benefit of the fundraising action is
\begin{equation}
MB^N=a\big[(1+\rho)L+(1-\rho)S\big],
\label{eq:privateMB}
\end{equation}
whereas its marginal contribution to aggregate mission output is
\begin{equation}
MB^M=2aL.
\label{eq:missionMB}
\end{equation}

\begin{proposition}[Lift--shift wedge]
For identical NGOs, the difference between the private marginal mission benefit of fundraising and its aggregate mission benefit is
\begin{equation}
MB^N-MB^M=a(1-\rho)(S-L).
\label{eq:wedge}
\end{equation}
\end{proposition}

The decomposition makes the externalities transparent. Shift is privately attractive because it raises own funding relative to rival funding. Lift creates a positive cross-organizational benefit because some newly attracted funding may accrue to the rival. Incomplete alignment overweights the first margin and underweights the second. Complete alignment eliminates both wedges simultaneously.

If marginal fundraising costs are increasing and the relevant solution is locally unique, equation~\eqref{eq:wedge} also characterizes the direction of the level distortion. Fundraising is excessive relative to the aggregate mission benchmark when $S>L$ and insufficient when $L>S$. The paper's main result, however, concerns the composition of fundraising rather than its total amount.

\section{Endogenous fundraising portfolios}
\label{sec:portfolio}

NGO $i$ can allocate fundraising effort across two technologies, $A$ and $B$. Total effort is
\begin{equation}
F_i=e_{iA}+e_{iB},
\end{equation}
and, for $F_i>0$, the share allocated to technology $A$ is
\begin{equation}
z_i=\frac{e_{iA}}{F_i}.
\end{equation}
A marginal unit of technology $k\in\{A,B\}$ generates lift $L_k$ and shift $S_k$. Its marginal mission value to the NGO is therefore
\begin{equation}
m_k(\rho)=a\big[(1+\rho)L_k+(1-\rho)S_k\big].
\label{eq:mk}
\end{equation}
Technology $A$ is \emph{more lift-intensive} than technology $B$ if
\begin{equation}
L_A-S_A>L_B-S_B.
\label{eq:liftorder}
\end{equation}
Neither technology needs to be pure lift or pure shift.

Fundraising uses scarce organizational capacity. Let its resource cost be
\begin{equation}
\Gamma(F_i,z_i)=C(F_i)+F_i h(z_i),
\label{eq:portfoliocost}
\end{equation}
where $C'>0$, $C''>0$, and $h''>0$. The first term captures the cost of total fundraising capacity. The second allows the resource requirement to depend on the portfolio, for example because technologies use specialized staff or communication channels. The cost functions do not depend on purpose alignment.

Conditional on the local funding responses, NGO $i$ chooses $F_i$ and $z_i$ to maximize
\begin{equation}
F_i\left[z_i m_A(\rho)+(1-z_i)m_B(\rho)\right]-C(F_i)-F_i h(z_i).
\label{eq:portfolioobjective}
\end{equation}
For an interior portfolio, the first-order condition for composition is
\begin{equation}
h'(z_i)=m_A(\rho)-m_B(\rho).
\label{eq:zFOC}
\end{equation}
The relative return satisfies
\begin{equation}
\frac{\partial}{\partial\rho}\left[m_A(\rho)-m_B(\rho)\right]
=a\left[(L_A-S_A)-(L_B-S_B)\right].
\label{eq:singlecrossing}
\end{equation}
The comparative static treats the local funding responses $(L_k,S_k)$ as characteristics of the fundraising technologies at the allocation under study. If these responses themselves vary with purpose alignment through other equilibrium adjustments, equation~\eqref{eq:singlecrossing} gives the direct effect; the same result holds for the total effect whenever the induced change in the lift--shift ranking does not overturn this single-crossing term.

\begin{proposition}[Purpose alignment changes the form of competition]
Suppose technology $A$ is more lift-intensive than technology $B$ as in equation~\eqref{eq:liftorder}. For an interior fundraising portfolio,
\begin{equation}
\frac{\partial z_i}{\partial\rho}
=\frac{a\left[(L_A-S_A)-(L_B-S_B)\right]}{h''(z_i)}>0.
\label{eq:dzdrho}
\end{equation}
Greater purpose alignment therefore shifts fundraising toward the relatively lift-intensive technology. The result is independent of the level-cost function $C(F_i)$ and does not require total fundraising to fall.
\label{prop:portfolio}
\end{proposition}

The result is a single-crossing property across fundraising technologies. Greater alignment raises the relative value of activities that expand funding available to the common purpose and lowers the relative value of activities that mainly redirect funding toward the focal NGO. The proposition is compositional. Alignment may raise or lower total fundraising because it changes both absolute marginal returns and the chosen portfolio. The unambiguous prediction concerns \emph{what type} of fundraising is used.

\section{Aggregate mission benchmark and coordination}
\label{sec:coordination}

For identical NGOs, aggregate mission value equals total mission output less real fundraising costs. The marginal aggregate mission value of one unit of fundraising technology $k$ is
\begin{equation}
w_k=2aL_k,
\label{eq:wk}
\end{equation}
because shift only reallocates funding between equally productive organizations. Conditional on total fundraising $F_i$, the mission-efficient portfolio $z_i^M$ satisfies
\begin{equation}
h'(z_i^M)=w_A-w_B=2a(L_A-L_B).
\label{eq:zmission}
\end{equation}
The privately chosen portfolio $z_i^N$ instead satisfies equation~\eqref{eq:zFOC}. Their relative-return wedge is
\begin{equation}
[w_A-w_B]-[m_A(\rho)-m_B(\rho)]
=a(1-\rho)\left[(L_A-S_A)-(L_B-S_B)\right].
\label{eq:portfoliowedge}
\end{equation}

\begin{proposition}[Coordination must target composition]
Suppose technology $A$ is more lift-intensive than technology $B$ and $\rho<1$. For an interior portfolio,
\begin{equation}
z_i^M>z_i^N.
\label{eq:zmzn}
\end{equation}
An incompletely aligned NGO therefore allocates too little fundraising capacity to the relatively lift-intensive technology from the perspective of aggregate mission value. A uniform per-unit tax or subsidy on all fundraising may change total effort but cannot correct this portfolio distortion.
\label{prop:coordination}
\end{proposition}

The second statement is an implementability result. A common per-unit charge applied to both technologies subtracts the same amount from both private marginal returns and therefore cancels from the composition condition. A blanket fundraising cap likewise operates on the level of effort rather than its relative allocation unless it directly constrains the mix. Reducing aggregate fundraising may therefore suppress valuable lift together with wasteful shift.

An instrument-specific correction may implement the mission benchmark. Let $p_k$ denote a per-unit subsidy, or an equivalent reduction in the resource cost, for technology $k$. Matching the private and mission marginal returns requires
\begin{equation}
p_k^*=a(1-\rho)(L_k-S_k).
\label{eq:pkstar}
\end{equation}
Hence
\begin{equation}
p_A^*-p_B^*=a(1-\rho)\left[(L_A-S_A)-(L_B-S_B)\right]>0.
\label{eq:relativepolicy}
\end{equation}
Equation~\eqref{eq:pkstar} is a corollary rather than the main policy result. Its role is to show how the same sufficient statistics that predict private fundraising composition also identify the relative correction needed to implement the mission benchmark. A fundraising federation, common platform, foundation, or public funder might create such relative incentives through matching support for cause-wide campaigns, shared infrastructure, differentiated fees, or restrictions on organization-specific solicitation.

This design question differs from asking whether a fundraising alliance is stable. \citet{AldashevMariniVerdier2014} study coalition formation and strategic interaction in fundraising coordination. Here the issue is how a coordinator concerned with aggregate mission output would alter relative incentives across fundraising activities when NGOs internalize rival mission output only partially. Experimental evidence that coordinated solicitation may itself increase total giving \citep{EckelGuneyUler2020} further illustrates why coordination cannot be evaluated solely by whether it reduces fundraising expenditure.

The benchmark is intentionally one of \emph{aggregate mission efficiency}, not complete social welfare. Donor utility and the opportunity cost of funds outside the NGO sector are not included. If donors give instrumentally because they value mission output, the benchmark captures the relevant organizational externality. If donors obtain warm glow from giving itself or additional utility from choosing a particular NGO, shift may carry direct donor surplus and the mission-efficient correction in equation~\eqref{eq:pkstar} need not coincide with the social-welfare correction. The composition result for NGO incentives remains valid; the policy interpretation then requires adding the corresponding donor-surplus term.

\section{A microfoundation: cause and organization promotion}
\label{sec:microfoundation}

A simple donation technology illustrates how the sufficient statistics arise. NGO $i$ chooses cause promotion $x_i$ and organization promotion $y_i$. Aggregate cause promotion is $X=x_1+x_2$. Total funding for the cause is $B(X)$, where $B'(X)>0$. Organization promotion determines NGO $i$'s funding share according to
\begin{equation}
s_i(y_i,y_j)=\frac{\exp(\tau y_i)}{\exp(\tau y_i)+\exp(\tau y_j)},
\qquad \tau>0,
\label{eq:share}
\end{equation}
and funding is
\begin{equation}
D_i=s_iB(X), \qquad D_j=(1-s_i)B(X).
\label{eq:funding}
\end{equation}
At a symmetric allocation, cause promotion changes both NGOs' funding by $B'(X)/2$. Hence
\begin{equation}
L_x=\frac{B'(X)}{2}, \qquad S_x=0.
\end{equation}
Organization promotion changes own and rival funding by equal amounts with opposite signs. Since $\partial s_i/\partial y_i=\tau/4$ at symmetry,
\begin{equation}
L_y=0, \qquad S_y=\frac{\tau B(X)}{4}.
\end{equation}
Cause promotion is therefore pure lift and organization promotion pure shift in this polar benchmark. Proposition~\ref{prop:portfolio} implies that greater purpose alignment shifts the fundraising portfolio toward cause promotion. No additional equilibrium comparative static is required.

At complete alignment, $\rho=1$, pure organization promotion has zero mission value for equally productive NGOs because it only relabels which organization controls a fixed pool. Cause promotion may remain valuable because it increases aggregate resources for the purpose. This benchmark clarifies why common purpose need not reduce fundraising overall: it changes the return to different forms of fundraising.

\section{Extensions}
\label{sec:extensions}

\subsection{Productivity differences and selection}

Suppose NGO $i$ and NGO $j$ have productivities $a_i$ and $a_j$. Consider a pure-shift activity that moves one marginal unit of funding from NGO $j$ to NGO $i$. The change in NGO $i$'s mission objective is proportional to
\begin{equation}
a_i-\rho_i a_j.
\label{eq:selectionvalue}
\end{equation}

\begin{proposition}[Selection]
Under complete purpose alignment, a pure-shift activity by NGO $i$ has positive mission value if and only if $a_i>a_j$. A less productive NGO has no mission-based incentive to divert a fixed funding pool from a more productive rival.
\label{prop:selection}
\end{proposition}

Purpose alignment therefore does not eliminate all organization-specific competition. It eliminates reallocation that leaves aggregate mission output unchanged or lowers it. Reallocation toward a more productive provider remains valuable. This result also limits the case for restricting organization promotion: relative promotion may be useful when it conveys information or directs funds toward more effective organizations. The present model represents that benefit through productivity differences and leaves donor learning for future work.

\subsection{Organization-specific value}

Purpose alignment need not imply indifference about which organization controls funding. Let $\omega_i\geq0$ denote organization-specific value per unit of funding controlled by NGO $i$. Its objective becomes
\begin{equation}
V_i=a_iD_i+\rho_i a_jD_j+\omega_iD_i-C_i.
\label{eq:omega}
\end{equation}
The term may represent autonomy, survival, reputation, future fundraising capacity, organization-specific learning, or private organizational rents. Some components may themselves have future mission value; others need not.

For a pure-shift activity, the marginal organizational return is proportional to
\begin{equation}
a_i+\omega_i-\rho_i a_j.
\end{equation}
Thus organization-centered competition may persist even when $\rho_i=1$ and $a_i=a_j$. The extension separates \emph{purpose alignment} from \emph{organization attachment}. Similar mission statements therefore provide only an imperfect empirical proxy for the cross-organizational valuation relevant to the model.

\section{Empirical and institutional implications}

The model suggests that fundraising expenditure should be decomposed by its funding effects rather than evaluated through a single overhead ratio. For a fundraising activity $k$, empirical estimates of its effects on own and rival funding identify
\begin{equation}
L_k=\frac12\left(\frac{\partial D_i}{\partial e_{ik}}+\frac{\partial D_j}{\partial e_{ik}}\right),
\qquad
S_k=\frac12\left(\frac{\partial D_i}{\partial e_{ik}}-\frac{\partial D_j}{\partial e_{ik}}\right).
\end{equation}
The literature already estimates related lift and displacement responses \citep{Meer2017,Schmitz2021,ScharfSmithOttoniWilhelm2022}. The additional empirical prediction is compositional: organizations with greater cross-organizational purpose alignment should allocate a larger share of their fundraising effort to technologies with higher $L_k-S_k$.

One possible empirical implementation would classify fundraising messages or channels by their category-level and organization-level content. Cause-wide awareness campaigns, common appeals, or shared platforms are natural candidates for relatively lift-intensive activities. Comparative claims, brand-building, and organization-specific donor acquisition are natural candidates for relatively shift-intensive activities, although the classification should ultimately rely on estimated funding responses rather than labels. Mission similarity measured from stated objectives may serve as one proxy for alignment, but the model predicts behavior as a function of cross-organizational valuation, not text similarity itself.

The institutional implication is equally specific. A federation, common platform, foundation, or public funder evaluating fundraising from the aggregate mission benchmark must distinguish activities that create additional resources for the common purpose from activities that mainly reallocate a fixed pool. Uniform limits on fundraising need not improve the mix. Relative support for shared donor acquisition, common campaign infrastructure, or cause-wide information moves the portfolio toward mission efficiency when those activities exhibit sufficiently high lift. Organization-specific promotion creates a mission loss through shift only to the extent that the reallocation does not generate offsetting productivity or informational gains. These statements concern the aggregate mission benchmark; a complete social-welfare prescription would additionally require donor surplus and the opportunity cost of funds outside the sector.

Public and foundation funding may also change subsequent fundraising composition. A grant may reduce the return to organization-specific solicitation, crowd out common donor acquisition, or alter the relative value of both. This is a market-wide additionality problem: the effect of one intervention depends on responses by competing organizations and on whether observed funding gains represent lift or displacement. \citet{JaagAdditionality2026} likewise defines additionality relative to a counterfactual that includes displacement and substitution.

\section{Conclusion}

Competition among NGOs with a common purpose differs from standard commercial rivalry because organizations may value output produced by their competitors. That feature has an ambiguous effect on how much they compete but a clear effect on how they compete.

The lift--shift decomposition identifies the relevant margins. Lift expands resources available to the common purpose. Shift reallocates existing resources between organizations. Incomplete purpose alignment causes an NGO to underweight lift that benefits its rival and to overweight shift that moves resources toward itself. When NGOs choose among fundraising technologies, greater alignment therefore shifts the portfolio toward the relatively lift-intensive technology.

The same lift--shift margins identify the institutional distortion. Relative to an aggregate mission benchmark, incompletely aligned NGOs choose portfolios that are too shift-intensive. Uniform fundraising regulation may change total effort while leaving this composition distortion intact. Coordination must instead alter relative incentives across technologies. The corresponding Pigouvian correction depends on lift minus shift, but the broader result is one of implementability: regulation of the level of fundraising and regulation of its composition are distinct problems.

The benchmark also clarifies its limits. Reallocation remains valuable when it directs funding toward a more productive organization, and organization-specific value may sustain rivalry even under complete purpose alignment. Donor surplus from choosing particular organizations may further change the welfare assessment of shift. These qualifications reinforce rather than weaken the central point: evaluating NGO competition requires asking what competitive effort does to the funding available for the common purpose, not merely how much organizations spend competing.

\appendix
\section{Proofs}

\subsection*{Proof of Proposition 1}
Using equation~\eqref{eq:LSresponses},
\begin{align*}
MB^N
&=a[(L+S)+\rho(L-S)]\\
&=a[(1+\rho)L+(1-\rho)S].
\end{align*}
Aggregate mission output changes by
\begin{equation*}
a[(L+S)+(L-S)]=2aL.
\end{equation*}
Subtracting yields equation~\eqref{eq:wedge}. \hfill$\square$

\subsection*{Proof of Proposition 2}
For an interior portfolio, equation~\eqref{eq:zFOC} holds. Differentiating with respect to $\rho$ gives
\begin{equation*}
h''(z_i)\frac{\partial z_i}{\partial\rho}
=a[(L_A-S_A)-(L_B-S_B)].
\end{equation*}
Strict convexity implies $h''(z_i)>0$. The lift-intensity ranking in equation~\eqref{eq:liftorder} makes the right-hand side positive, which proves equation~\eqref{eq:dzdrho}. \hfill$\square$

\subsection*{Proof of Proposition 3}
Equation~\eqref{eq:portfoliowedge} is positive when $A$ is more lift-intensive than $B$ and $\rho<1$. Since $h''>0$, the composition condition is strictly increasing in $z_i$, so $z_i^M>z_i^N$. A common per-unit charge applied to both technologies subtracts the same amount from both private marginal returns and cancels from their difference. It therefore does not change the portfolio first-order condition. \hfill$\square$

\subsection*{Proof of Proposition 4}
A pure-shift activity transfers a marginal unit of funding from NGO $j$ to NGO $i$. NGO $i$ gains $a_i$ units of own mission output and loses $\rho_i a_j$ units of valued rival output. Its marginal mission return is therefore proportional to $a_i-\rho_i a_j$. At $\rho_i=1$, the expression is positive if and only if $a_i>a_j$. \hfill$\square$

\clearpage
\section*{Declaration of generative AI and AI-assisted technologies in the writing process}
During the preparation of this work the author used ChatGPT (OpenAI) to review theoretical derivations, refine manuscript structure, and improve readability. The author independently verified the derivations and reviewed and edited the AI-assisted output as needed, and takes full responsibility for the content of the published article.


\begin{thebibliography}{99}

\bibitem[Aldashev and Verdier(2009)]{AldashevVerdier2009}
Aldashev, G., and T. Verdier (2009).
When NGOs go global: Competition on international markets for development donations.
\emph{Journal of International Economics} 79(2), 198--210.
https://doi.org/10.1016/j.jinteco.2009.07.007.

\bibitem[Aldashev and Verdier(2010)]{AldashevVerdier2010}
Aldashev, G., and T. Verdier (2010).
Goodwill bazaar: NGO competition and giving to development.
\emph{Journal of Development Economics} 91(1), 48--63.
https://doi.org/10.1016/j.jdeveco.2008.11.010.

\bibitem[Aldashev et~al.(2014)Aldashev, Marini, and Verdier]{AldashevMariniVerdier2014}
Aldashev, G., M. Marini, and T. Verdier (2014).
Brothers in alms? Coordination between nonprofits on markets for donations.
\emph{Journal of Public Economics} 117, 182--200.
https://doi.org/10.1016/j.jpubeco.2014.04.009.

\bibitem[Aldashev et~al.(2020)Aldashev, Marini, and Verdier]{AldashevMariniVerdier2020}
Aldashev, G., M. Marini, and T. Verdier (2020).
Samaritan bundles: Fundraising competition and inefficient clustering in NGO projects.
\emph{Economic Journal} 130(630), 1541--1582.
https://doi.org/10.1093/ej/ueaa031.

\bibitem[Besley and Ghatak(2005)]{BesleyGhatak2005}
Besley, T., and M. Ghatak (2005).
Competition and incentives with motivated agents.
\emph{American Economic Review} 95(3), 616--636.
https://doi.org/10.1257/0002828054201413.

\bibitem[Bilodeau and Slivinski(1997)]{BilodeauSlivinski1997}
Bilodeau, M., and A. Slivinski (1997).
Rival charities.
\emph{Journal of Public Economics} 66(3), 449--467.
https://doi.org/10.1016/S0047-2727(97)00046-7.

\bibitem[Burani and Manna(2026)]{BuraniManna2026}
Burani, N., and E. Manna (2026).
Bad NGOs? Competition in the market for donations and workers' misconduct.
\emph{Journal of Industrial Economics} 74(3), 367--383.
https://doi.org/10.1111/joie.70027.

\bibitem[Castaneda et~al.(2008)Castaneda, Garen, and Thornton]{CastanedaGarenThornton2008}
Castaneda, M. A., J. Garen, and J. Thornton (2008).
Competition, contractibility, and the market for donors to nonprofits.
\emph{Journal of Law, Economics, and Organization} 24(1), 215--246.
https://doi.org/10.1093/jleo/ewm036.

\bibitem[Chau and Huysentruyt(2006)]{ChauHuysentruyt2006}
Chau, N. H., and M. Huysentruyt (2006).
Nonprofits and public good provision: A contest based on compromises.
\emph{European Economic Review} 50(8), 1909--1935.
https://doi.org/10.1016/j.euroecorev.2005.11.006.

\bibitem[Ek(2017)]{Ek2017}
Ek, C. (2017).
Some causes are more equal than others? The effect of similarity on substitution in charitable giving.
\emph{Journal of Economic Behavior \& Organization} 136, 45--62.
https://doi.org/10.1016/j.jebo.2017.01.007.

\bibitem[Eckel et~al.(2020)Eckel, Guney, and Uler]{EckelGuneyUler2020}
Eckel, C. C., B. Guney, and N. Uler (2020).
Independent vs. coordinated fundraising: Understanding the role of information.
\emph{European Economic Review} 127, 103476.
https://doi.org/10.1016/j.euroecorev.2020.103476.

\bibitem[Gayle and Harrison(2026)]{GayleHarrison2026}
Gayle, P. G., and T. D. Harrison (2026).
Competition and nonprofit's strategic responses: Evidence from fundraising in donative markets.
\emph{Review of Economics and Statistics}, forthcoming.
https://doi.org/10.1162/REST.a.1778.

\bibitem[Gallier et~al.(2023)Gallier, Goeschl, Kesternich, Lohse, Reif, and R\"omer]{GallierEtAl2023}
Gallier, C., T. Goeschl, M. Kesternich, J. Lohse, C. Reif, and D. R\"omer (2023).
Inter-charity competition under spatial differentiation: Sorting, crowding, and spillovers.
\emph{Journal of Economic Behavior \& Organization} 216, 457--468.
https://doi.org/10.1016/j.jebo.2023.10.013.

\bibitem[Glaeser and Shleifer(2001)]{GlaeserShleifer2001}
Glaeser, E. L., and A. Shleifer (2001).
Not-for-profit entrepreneurs.
\emph{Journal of Public Economics} 81(1), 99--115.
https://doi.org/10.1016/S0047-2727(00)00130-4.

\bibitem[Hagen and Ratts\o(2024)]{HagenRattso2024}
Hagen, R. J., and J. Ratts\o{} (2024).
Strategic interaction in the market for charitable donations: The role of public funding.
\emph{Journal of Public Economic Theory} 26(1), e12681.
https://doi.org/10.1111/jpet.12681.

\bibitem[Jaag(2026a)]{JaagAdditionality2026}
Jaag, C. (2026a).
On additionality: Output, employment, and diversity in impact investing.
Unpublished manuscript.
https://doi.org/10.13140/RG.2.2.24403.54563/1.

\bibitem[Jaag(2026b)]{JaagGrantImpactLoan2026}
Jaag, C. (2026b).
Grant or impact loan? Impact capital under scarcity.
Unpublished manuscript.
https://doi.org/10.13140/RG.2.2.14730.43202/1.

\bibitem[Jaag(2026c)]{JaagCatalytic2026}
Jaag, C. (2026c).
Catalytic capital and impact governance: A theory of commercial crowding-in.
Unpublished manuscript.
https://doi.org/10.13140/RG.2.2.34653.37609/1.

\bibitem[Jaag(2026d)]{JaagFutureIntervention2026}
Jaag, C. (2026d).
When future intervention weakens mission: Organizational commitment under future contracting.
Unpublished manuscript.
https://doi.org/10.13140/RG.2.2.14861.50407/2.

\bibitem[Krasteva and Yildirim(2016)]{KrastevaYildirim2016}
Krasteva, S., and H. Yildirim (2016).
Information, competition, and the quality of charities.
\emph{Journal of Public Economics} 144, 64--77.
https://doi.org/10.1016/j.jpubeco.2016.10.008.

\bibitem[Ly and Mason(2012)]{LyMason2012}
Ly, P., and G. Mason (2012).
Competition between microfinance NGOs: Evidence from Kiva.
\emph{World Development} 40(3), 643--655.
https://doi.org/10.1016/j.worlddev.2011.09.009.

\bibitem[Meer(2017)]{Meer2017}
Meer, J. (2017).
Does fundraising create new giving?
\emph{Journal of Public Economics} 145, 82--93.
https://doi.org/10.1016/j.jpubeco.2016.11.009.

\bibitem[Philipson and Posner(2009)]{PhilipsonPosner2009}
Philipson, T. J., and R. A. Posner (2009).
Antitrust in the not-for-profit sector.
\emph{Journal of Law and Economics} 52(1), 1--18.
https://doi.org/10.1086/589704.

\bibitem[Rose-Ackerman(1982)]{RoseAckerman1982}
Rose-Ackerman, S. (1982).
Charitable giving and ``excessive'' fundraising.
\emph{Quarterly Journal of Economics} 97(2), 193--212.
https://doi.org/10.2307/1880754.

\bibitem[Scharf(2014)]{Scharf2014}
Scharf, K. (2014).
Impure prosocial motivation in charity provision: Warm-glow charities and implications for public funding.
\emph{Journal of Public Economics} 114, 50--57.
https://doi.org/10.1016/j.jpubeco.2013.10.002.

\bibitem[Scharf et~al.(2022)Scharf, Smith, and Ottoni-Wilhelm]{ScharfSmithOttoniWilhelm2022}
Scharf, K., S. Smith, and M. Ottoni-Wilhelm (2022).
Lift and shift: The effect of fundraising interventions in charity space and time.
\emph{American Economic Journal: Economic Policy} 14(3), 296--321.
https://doi.org/10.1257/pol.20180679.

\bibitem[Schmitz(2021)]{Schmitz2021}
Schmitz, J. (2021).
Is charitable giving a zero-sum game? The effect of competition between charities on giving behavior.
\emph{Management Science} 67(10), 6333--6349.
https://doi.org/10.1287/mnsc.2020.3809.

\end{thebibliography}
\end{document}